\documentclass[sigconf]{acmart}

\AtBeginDocument{%
  \providecommand\BibTeX{{%
    Bib\TeX}}}

\pdfoutput=1
\usepackage{blindtext}
\usepackage{soul}
\usepackage{pgfplots}
\usetikzlibrary{decorations.pathmorphing, patterns}
\pgfplotsset{compat=1.17}
\usepackage{dsfont}
\usepackage{enumitem}
\usepackage{algcompatible}
\usepackage{times}
\usepackage{booktabs} 
\usepackage{mdframed}
\usepackage{framed}
\usepgfplotslibrary{groupplots}
\usepackage{hyphenat}
\usepackage{fnpct}
\usepackage{algorithm}
\usepackage{algpseudocode}
\usepackage{amsmath,epsfig}
\usepackage{amsthm}
\usepackage{setspace}
\usepackage{enumitem}
\usepackage{multirow}
\usepackage{hhline}
\usepackage{caption}
\usepackage{subcaption}
\usepackage{graphicx}
\usepackage{wrapfig}
\usepackage{amsmath,epsfig}
\usepackage{amsthm}

\setcopyright{rightsretained}
\usepackage{tikz}
\usepackage{balance}

\usepackage{csquotes}
\usepackage{enumitem}
\usepackage{tablefootnote}
\usepackage{fnpct}
\usepackage{bbm}
\usepackage{lipsum}
\usepackage{setspace}
\usepackage{adjustbox}

\makeatletter
\def\thm@space@setup{\thm@preskip=0pt
\thm@postskip=0pt}
\makeatother

\usepackage{array}
\usepackage{makecell}

\usepackage{url}
\usepackage{balance}

\usepackage{longtable}
\usepackage{mdframed}
\usepackage{colortbl}

\usepackage[hang,flushmargin]{footmisc}

\newcommand*\wrapletters[1]{\wr@pletters#1\@nil}
\def\wr@pletters#1#2\@nil{#1\allowbreak\if&#2&\else\wr@pletters#2\@nil\fi}
\usepackage{tikz}
\usetikzlibrary{automata,matrix,shapes,arrows,positioning,chains,calc}
\usetikzlibrary{snakes}
\usetikzlibrary{arrows,scopes}
\usetikzlibrary{positioning,chains,fit,shapes,calc}
\usepackage{caption}
\usepackage{subcaption}
\usepackage{amsmath}
\usepackage{lineno}
\usepackage[inkscapelatex=false]{svg}
\definecolor{lightyellow}{HTML}{ABEBC6}
\sethlcolor{lightyellow}

\definecolor{myblue}{HTML}
{BF40BF}

\def\BibTeX{{\rm B\kern-.05em{\sc i\kern-.025em b}\kern-.08em
    T\kern-.1667em\lower.7ex\hbox{E}\kern-.125emX}}

\setcopyright{acmlicensed}
\copyrightyear{2018}
\acmYear{2018}
\acmDOI{XXXXXXX.XXXXXXX}
\acmConference[Conference acronym 'XX]{Make sure to enter the correct
  conference title from your rights confirmation email}{June 03--05,
  2018}{Woodstock, NY}
\acmISBN{978-1-4503-XXXX-X/18/06}

\begin{document}

\title{Empirical Analysis of EIP-3675: Miner Dynamics, Transaction Fees, and Transaction Time}


\author{Umesh Bhatt}
\affiliation{%
  \department{Data Science and Applications}
  \institution{Indian Institute of Technology, Madras}
  \city{Chennai}
  \country{India}}
\email{umesh.bhatt2002@gmail.com}

\author{Sarvesh Pandey}
\affiliation{%
  \department{Computer Science - MMV}
  \institution{Banaras Hindu University}
  \city{Varanasi}
  \country{India}}
\email{sarveshpandey@bhu.ac.in}

\begin{abstract}
The Ethereum Improvement Proposal 3675 (EIP-3675) marks a significant shift, transitioning from a Proof of Work (PoW) to a Proof of Stake (PoS) consensus mechanism. This transition resulted in a staggering 99.95\% decrease in energy consumption. However, the transition prompts two critical questions: (1). How does EIP-3675 affect miners' dynamics? and (2). How do users determine priority fees, considering that paying too little may cause delays or non-inclusion, yet paying too much wastes money with little to no benefits? To address the first question, we present a comprehensive empirical study examining EIP-3675's effect on miner dynamics (i.e., miner participation, distribution, and the degree of randomness in miner selection). Our findings reveal that the transition has encouraged broader participation of miners in block append operation, resulting in a larger pool of unique miners ($\approx50\times$ PoW), and the change in miner distribution with the increased number of unique \textit{small category} miners ($\approx60\times$ PoW). However, there is an unintended consequence: a reduction in the miner selection randomness, which signifies the negative impact of the transition to PoS-Ethereum on network decentralization. Regarding the second question, we employed regression-based machine learning models; the Gradient Boosting Regressor performed best in predicting priority fees, while the K-Neighbours Regressor was worst.
\end{abstract}



\maketitle

\section{Introduction}\label{intro}
Blockchain facilitates decentralized and transparent storage of transactions in a shared immutable ledger; notable examples are Bitcoin~\cite{Bitcoin} and Ethereum~\cite{Ethereum}. In contrast to conventional centralized databases (where a limited number of well-defined nodes primarily perform append/write operations), blockchains facilitate consensus-driven write operations (or block append operations) and disperse transactional data across a network of peers. Using cryptographic techniques, including hashing, a set of transactions is wrapped into a "block" and appended to a chain of previously validated blocks, forming an unalterable history~\cite{DBLP:journals/corr/abs-1906-11078}. Maintaining a decentralized ledger eliminates the need for intermediaries and thus lowers the possibility of fraud while allowing the trustless execution of transactions between peers. 
     
Critical to the functioning of any blockchain is the consensus mechanism, which ensures agreement among nodes on the present state of the shared ledger~\cite{DBLP:journals/csur/00310J23,DBLP:conf/icde/KantPS22}. Proof-based consensus algorithms have been designed for permissionless blockchains in the past. Proof of Work (PoW)~\cite{nakamoto2008bitcoin} is the first proof-based algorithm; several variants of PoW have been proposed in the last decade, e.g., Proof of Stake (PoS)~\cite{wood2014ethereum}, and Proof of Authority (PoA)~\cite{PoA}. PoW consensus has been at the core of Bitcoin since its inception in \emph{2009} to facilitate block append operation~\cite{DBLP:journals/corr/abs-1906-11078}. Furthermore, Ethereum also adopted the PoW consensus mechanism~\cite{nakamoto2008bitcoin} from its inception in \emph{2013} until the \emph{Paris upgrade (Sept 15, 2022)}.  However, a transformative moment in Ethereum's evolution occurred with the \emph{Paris upgrade}, introducing PoS through \emph{EIP-3675}~\cite{EthHistory}. Interstingly, Bitcoin -- Ethereum's predecessor and the first decentralized cryptocurrency -- has no plans to change its PoW algorithm because of security concerns, e.g., 33\% attack~\cite{33Etc, 33Eth}.

PoW requires miners to spend computational resources to solve a computation puzzle; a miner that solves it first is incentivized to append the newly mined block to the shared ledger, while PoS requires miners to stake cryptocurrency (e.g., a minimum of 32 Ether has to be staked by validators to participate in mining in the case of Ethereum) as collateral to create, validate, and append new blocks. `Miner' and `Proposer/Validator' refer to block producers in Bitcoin and Ethereum, respectively. Moreover, \emph{`Paris upgrade'} and \emph{`EIP-3675'} are interchangeably used in this paper. 

The Ethereum community's switch to PoS is a decision taken with the intent to make the consensus process energy-efficient without sacrificing the blockchain's core characteristics. The Paris upgrade~\cite{EthHistory} is facilitated by introducing the \emph{RANDAO} algorithm~\cite{BlockProposal}, which randomly selects a proposer from a set of validators based on the Ether they hold and are willing to stake. The more Ether a validator stakes, the higher their chances of getting selected as block-proposer and proposing a new block.  This upgrade also required the active role of the beacon chain (initiated on \emph{Dec 1, 2020}) in storing consensus data in slots, including attestations, withdrawals, validator data, and empty execution payloads. However, as designed, after the Paris upgrade, the beacon chain stores transaction-related data in the execution payload~\cite{ConsensusSpecs}. 

\subsection {Use Case}  
\begin{figure}[htbp]
    \centering
    \includegraphics[width=0.8\columnwidth]{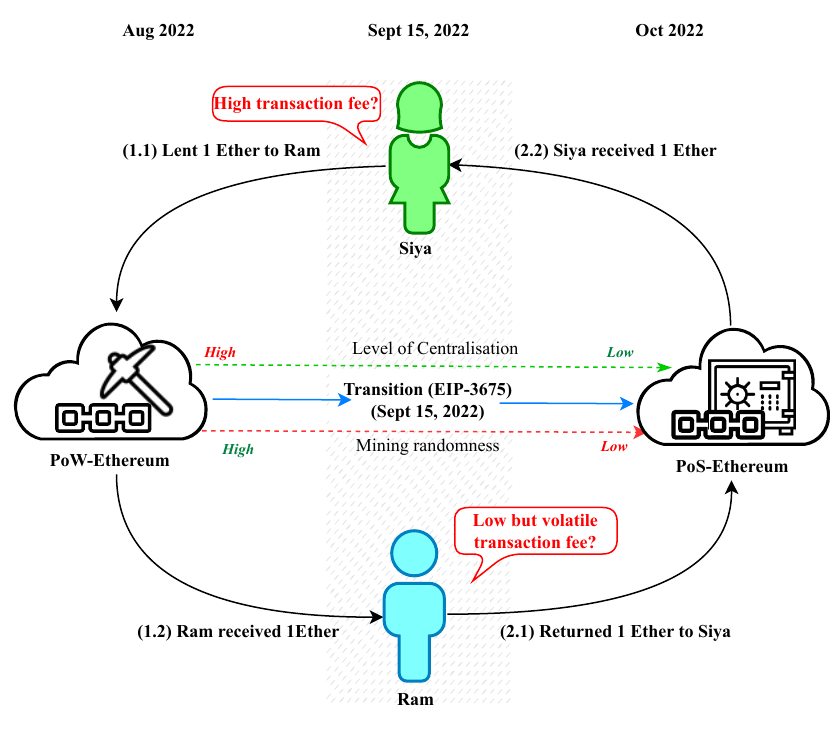}
    \caption{Problem Statement}
    \label{fig:usecase}
\end{figure}

Let us illustrate a practical scenario involving transactions between \texttt{\textsl{Ram}} and \texttt{\textsl{Siya}} (graduate students), showing the impact of \emph{EIP-3675} on users submitting transactions (e.g., send/receive operation) (see Fig.~\ref{fig:usecase}). \texttt{\textsl{Ram}} is exploring blockchain platforms, has developed a blockchain algorithm, and needs to test it on various blockchain frameworks like Ethereum, Bitcoin, and Hyperledger Fabric. However, because of his laptop's limited capabilities, he realized that he could not fully sync Ethereum and Bitcoin and thus required one additional high-performance system. He sought to purchase a workstation but lacked sufficient funds, having only 2 Ether while the workstation cost 3 Ether. 

\texttt{\textsl{Ram}} turned to his friend \texttt{\textsl{Siya}}, who lent him 1 Ether in \textbf{\textit{Aug 2022}} with the understanding that he would repay her within two months. \texttt{\textsl{Siya}} had to pay a high transaction fee (\emph{txnFee}) for adding this transaction to the Ethereum ledger due to Ethereum's PoW protocol but proceeded with the transaction. \texttt{\textsl{Ram}} purchased the workstation with \texttt{\textsl{Siya}}'s help and completed the research task. In \textbf{\textit{Sept 2022}}, Ethereum transitioned to PoS, which altered mining dynamics, reducing energy consumption but introducing fluctuations in \emph{txnFee}. Then, in \textbf{\textit{Oct 2022}}, \texttt{\textsl{Ram}} decides to repay \texttt{\textsl{Siya}}. He is happy that the \emph{txnFee} he needs to pay to include his transaction in the Ethereum ledger is lesser after the \emph{EIP-3675} improvement. However, he needs to carefully determine \emph{txnFee} to avoid delay or unnecessary expense, as the priority fee (a \emph{txnFee} component) is volatile and can only offer benefits up to a certain extent. This use case highlights the real-world implications of our research, showing how the shift to PoS has impacted \texttt{\textsl{Siya}} and \texttt{\textsl{Ram}}, who transact in Ether.

The transaction fee mechanism (TFM) in Ethereum is dynamic and multifaceted, governing the incentives for proposers and validators and, thus, shaping the economic landscape of the network~\cite{DBLP:conf/soda/ChungS23, DBLP:journals/corr/abs-2012-00854}. In the case of \emph{PoW-Ethereum}, \emph{txnFee} -- derived from base fee (\emph{baseFee}) and priority fee (\emph{priorityFee}) -- was influenced by numerous factors, e.g., network congestion, the availability of block space, and the competition among miners to include transactions in the next block. Driven by the prospect of earning more, Miners prioritized transactions based on \emph{txnFee}. 

In \emph{PoS-Ethereum}, Proposers append blocks~\cite{PoW-vs-PoS}; it has the following positives: significantly reduced energy usage (by about 99.95\%), faster block confirmations, and lower \emph{txnFee}~\cite{EthEnergy}. With the substantial decrease in the cost of the consensus mechanism with \emph{PoS-Ethereum}, which no longer requires expensive puzzle-solving computations, proposers' average \emph{priorityFee} on block append operations has reduced considerably. However, the \emph{priorityFee} factor has become more volatile. Therefore, it is crucial to understand how this transition impacts \emph{priorityFee}.  

Exploring the miner centralization in \emph{PoS-Ethereum} involves analyzing unique miner participation, miner distribution  (small, medium, and large), and miner selection randomness. It is crucial for a comprehensive analysis to consider the implications for \textit{\textbf{small-scale}} validators and how the network incentivizes their participation. The gains from the reduced computational work have altered the competitive dynamics among validators (block producers), influencing the fee structure in various ways. It will also be interesting to explore how the stakes held by validators correlate with the transaction fees they set. Does a higher stake translate to higher fees? Or does the \emph{PoS-Ethereum} achieve the intended trade-off on a fair fee-setting mechanism? \emph{PoS-Ethereum} also fuelled concerns about potential centralization due to wealth concentration. The rich-get-richer debate revolves around whether wealthier miners with more stakes gain higher staking rewards, potentially amplifying income inequality in the future~\cite{DBLP:conf/sigmod/HuangTCLX21}. The question is whether validators with higher stakes may gain disproportionate rewards, potentially leading to wealth concentration and centralization within the mining ecosystem.

We have worked on two critical aspects of the impact of \emph{EIP-3675}: \textbf{(a).} Miner dynamics analysis (\emph{PoW-Ethereum} and \emph{PoS-Ethereum}), i.e., examining the effects of \emph{EIP-3675} on miner participation level, miner distribution, and the degree of randomness in miner selection. The insights drawn will have long-term practical implications for the Ethereum community in understanding the potential change in the mining landscape. \textbf{(b).} Machine learning models to predict \emph{txnFee} and \emph{txnTime} for the \emph{PoS-Ethereum} setting. This is particularly relevant for end users, who need to make informed decisions about \emph{txnFee} in the \emph{PoS-Ethereum} (refer to Fig.~\ref{fig:usecase} for better clarity).
    
\textbf{\emph{Organization.}} The research paper is structured as follows: \emph{Section~\ref{back}} introduces
the TFM and details of the \emph{EIP-3675} upgrade. \emph{Section~\ref{data}} discusses data sources utilized and dataset generation steps. \emph{Section~\ref{emp}} details the research methodology and presents the results of the empirical analysis. \emph{Section~\ref{rel_work}} reviews related work, providing a context for existing analytical studies on Ethereum and TFM. \emph{Section~\ref{conclusion}} consolidates the findings and discusses their significance.

\begin{table*}[h]
\footnotesize
\centering
\caption{Notations related to EIP-3675}
\label{tab:notations}
\begin{tabular}{|l|l|}
\hline
\textbf{Concept} & \textbf{Description}                                                                                                  \\ \hline
PoW-Ethereum     & Ethereum during its Proof of Work phase, where miners compete to validate transactions through complex puzzles.       \\ \hline
PoS-Ethereum     & Ethereum during its Proof of Stake phase, where validators create and validate blocks based on staked cryptocurrency. \\ \hline
Miner            & Nodes in Proof of Work that solve cryptographic puzzles to add blocks.                                                \\ \hline
Validator        & A virtual entity in Proof of Stake validating and creating blocks based on staked cryptocurrency.                     \\ \hline
Proposer         & Pseudorandomly selected validator to append blocks in Proof of Stake for given slots.                                 \\ \hline
EIPs             & Ethereum Improvement Proposals are the standards defining new features for Ethereum.                                  \\ \hline
gas              & Unit measuring computational effort for operations on Ethereum.                                                       \\ \hline
transactionFee   & Cost paid in gas for executing transactions on Ethereum, ensuring network resources and security.                     \\ \hline
\end{tabular}
\end{table*}

\section{Background}\label{back}
This section presents a brief overview of TFM in Ethereum (see Section~\ref{TFM}) followed by a discussion on the impact of EIP-3675 on TFM (see section~\ref{impact}). 

\subsection{Transaction Fee Mechanism}\label{TFM}
The TFM in Ethereum represents how \emph{txnFee} is managed within the network. Ethereum Improvement Proposal 1559 (\emph{EIP-1559})~\cite{EIP-1559} introduces a dynamic fee structure, which is adjusted based on block gas usage. Users can bid on transactions using parameters such as max priority fee per gas and max fee per gas, ensuring backward compatibility while providing flexibility and fairness in transaction pricing. Crucially, \emph{EIP-1559} burns the \emph{baseFee}, removing it from circulation while remitting \emph{priorityFee} to miners as rewards, incentivizing transaction inclusion and discouraging empty block mining. This mechanism enhances efficiency and user experience in fee management.

\subsection{Impact of \emph{EIP-3675} on TFM}\label{impact}
\emph{EIP-3675} facilitates Ethereum's transition from PoW to PoS consensus mechanism. It has significant implications for the TFM within the Ethereum network. With the adoption of PoS, the costly puzzle-solving computations inherent in PoW are eliminated, reducing the average \emph{priorityFee} required for block append operations. This reduction reflects a more cost-effective approach to achieving consensus, which aligns with Ethereum's goal of improving scalability and sustainability. However, the transition also introduced greater volatility in the \emph{priorityFee} factor due to changes in miner dynamics.

\section{Data}\label{data}

\subsection{Data Sources}
We collect data from three different sources to generate a dataset. \textbf{\textit{First}}, we synced the \emph{Ethereum full node} in our local system using the execution client Geth~\cite{Geth} and consensus client Prysm~\cite{Prysm}. The local \emph{Ethereum full node} was a primary transaction-specific and consensus-related data source. It included variables such as block number, miner public key, transaction value, timestamps, slot number, and validator data. \textbf{\textit{Second}}, we fetched additional transaction-specific data from \emph{Google BigQuery}~\cite{GBigquerry}, e.g., gas price. \textit{\textbf{Third}}, we fetched additional consensus-specific data from \emph{Beaconcha.in}~\cite{beaconCha.in}, e.g., the number of active validators. 

\begin{figure}[htbp]
    \centering
    \includegraphics[width=\columnwidth]{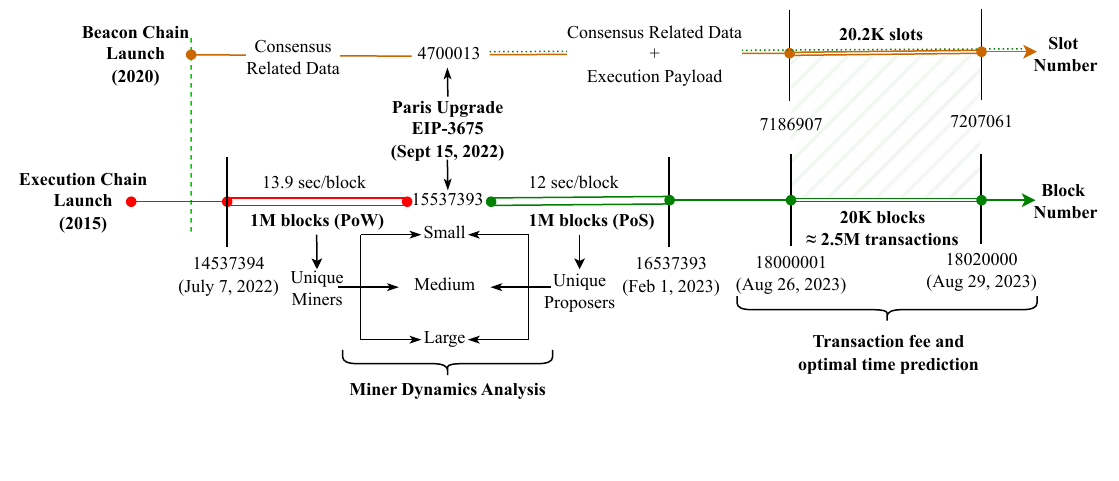}
    \caption{Data Sources and Block Ranges Considered}
    \label{fig:proposedWork}
\end{figure}

We sliced the data collected for two different time ranges, i.e., from \emph{July 7, 2022}, to \emph{Feb 1, 2023}, and from \emph{Aug 26, 2023}, to \emph{Aug 29, 2023}, for miner dynamics analysis and transaction fee and time prediction, respectively (see Fig.~\ref{fig:proposedWork}). For miner dynamics analysis, the last 1 million blocks were selected for \emph{PoW-Ethereum} (from block number \emph{14537394} to \emph{15537393}), and the initial 1 million blocks were chosen for \emph{PoS-Ethereum} (from block number \emph{15537394} to \emph{16537393}), providing a comprehensive temporal scope for understanding the impact of Ethereum's \emph{Paris upgrade} on miner behavior. We used data comprising more than 2.5 million transactions(from block number \emph{18000001} to \emph{18020000}) to train the regression-based machine-learning models to predict \emph{txnFee} and \emph{txnTime}. For better comprehension and clarity, we have described the common notations related to \emph{EIP-3675} used in this paper in Table~\ref{tab:notations}. 

\subsection{Miner Data}
We obtain block header data from the \emph{Ethereum full node}~\cite{Geth, Prysm}, including the miner's public key for the last 1M \emph{PoW-Ethereum} blocks and the initial 1M \emph{PoS-Ethereum} blocks. For simplicity,  we exported this data to the \emph{JSON} format. Subsequently, we partition the data across three distinct volumes, each containing 100K, 500K, and 1M blocks. We extract the unique miner's public keys and the blocks they mined for each volume from the extracted header data. 

\subsection{Transaction Data}
We obtain transaction-related data, including gas used (\emph{gasUsed}), gas price (\emph{gasPrice}), and base fee per gas (\emph{baseFeePerGas}) for each transaction from the \emph{Ethereum full node}~\cite{Geth, Prysm} and \emph{Google BigQuery}~\cite{GBigquerry}. To extract relevant features for our machine learning model, we employed the following key transformations: 
\emph{baseFee} is calculated as the product of \emph{gasUsed} and the \emph{baseFeePerGas} (see Eq.~\ref{eqn:base}),
        \begin{equation} \label{eqn:base}
        baseFee = gasUsed*baseFeePerGas
        \end{equation}
\emph{txnFee} is derived from the product of \emph{gasUsed} and \emph{gasPrice} (see Eq.~\ref{eqn:txnFee}),
        \begin{equation} \label{eqn:txnFee}
        txnFee = gasUsed*gasPrice
        \end{equation}
and \emph{priorityFee} represents the surplus transaction cost beyond the \emph{baseFee} (see Eq.~\ref{eqn:priority}).
        \begin{equation} \label{eqn:priority}
        priorityFee = txnFee - baseFee
        \end{equation}

We extract consensus-related data, including total votes and active validators from \emph{Ethereum full node}~\cite{Geth, Prysm} and \emph{Beaconcha.in}~\cite{beaconCha.in}, and exported it to the \emph{JSON} format for simplicity. 
Then, to combine transaction data with consensus data, we introduced the Block Slot Mapping (\textsc{BSMap++}) algorithm (see Algorithm~\ref{alg:BSMap++}). \textsc{BSMap++} serves as a strategic solution to establish a reliable mapping between main chain blocks and beacon slots, which enhances the overall integrity and coherence of the dataset.

\begin{algorithm}[!t]
\caption{\textsc{BSMap}}
\label{alg:BSMap}
\footnotesize
\raggedright
\textbf{Input:} Starting and ending slot numbers for beacon chain and block number of main chain block;\\
\textbf{Output:} Slot number of beacon block corresponding to the main chain's block number;
\begin{algorithmic}[1]
\Procedure{BSMap}{$beaconStart, beaconEnd, mainChainBlockNum$}
    \State $m \gets beaconStart$
    
    \State $n \gets beaconEnd$ 
    \State $mcb \gets mainChainBlockNum$
    \While{$m \leq n$}
        \State $beaconSlot \gets  \lfloor(m+n)/2  \rfloor$
        \State $beaconChainBlockNum \gets \emph{fetchBlockNumber(beaconSlot)}$
        \If{$mcb = beaconChainBlockNum$}
            \State \textbf{return} $beaconSlot$
        \ElsIf{$mcb > beaconChainBlockNum$}
            \State $m \gets beaconSlot+1$
        \Else
            \State $n \gets beaconSlot-1$
        \EndIf
    \EndWhile
\State \textbf{return} $-1$
\EndProcedure 

\end{algorithmic}
\end{algorithm}

\begin{algorithm}
\caption{\textsc{BSMap++}}
\label{alg:BSMap++}
\footnotesize
\raggedright
\textbf{Input:} Starting and ending slot numbers for beacon chain, block number of main chain block and number of partitions;\\
\textbf{Output:} Slot number of beacon block corresponding the main chain's block number;
\begin{algorithmic}[1]
\Procedure{BSMap}{$beaconStart, beaconEnd, mainChainBlockNum, k$}
\State $m \gets beaconStart$
\State $n \gets beaconEnd$
\State $mcb \gets mainChainBlockNum$
\If{$k \leq 1$}
    \For{$beaconSlot \gets m$ \textbf{to} $n$}
    \State $beaconChainBlockNum \gets \emph{fetchBlockNumber(beaconSlot)}$
        \If{$mcb = beaconChainBlockNum$}
            \State \textbf{return} $beaconSlot$
        \EndIf
    \EndFor
    \State \Return $-1$
\EndIf
\While{$m \leq n$}
    \State $binSize \gets \left\lfloor \frac{n - m}{k} \right\rfloor$
    \State $beaconSlots \gets \{m + i \cdot binSize \mid i \in \{0, 1, \ldots, k-1\}\} \cup \{n\}$
    \State $bcbs \gets \{\emph{fetchBlockNumber}(slot) \mid slot \in beaconSlots\}$
    \State $found \gets \textbf{False}$
    \For{$i \gets 1$ \textbf{to} $|\text{beaconSlots}|-1$}
        \If{$\text{bcbs}[i] = mcb$}
            \State \Return $\text{beaconSlots}[i]$
        \EndIf
        \If{$\text{bcbs}[i] > mcb > \text{bcbs}[i-1]$}
            \State $n \gets \text{beaconSlots}[i] - 1$
            \State $m \gets \text{beaconSlots}[i - 1]$
            \State $found \gets \textbf{True}$
            \State \textbf{break}
        \EndIf
    \EndFor
    \If{\textbf{not} $found$}
        \State \textbf{break}
    \EndIf
\EndWhile
\State \Return $-1$
\EndProcedure 
\end{algorithmic}
\end{algorithm}

\begin{algorithm}
\caption{\textsc{ComputeOptimumK}}
\label{alg:findOptimumK}
\footnotesize
\raggedright
\textbf{Input:} Starting and ending block numbers for main chain, starting and ending slot numbers for beacon chain, the maximum number of partitions;\\
\textbf{Output:} Optimum number of partitions;
\begin{algorithmic}[1]
\Procedure{ComputeOptimumK}{$mainChainStart, mainChainEnd, beaconStart, beaconEnd, n$}
    \State $test \gets \text{RandomSample(mainChainStart, mainChainEnd, 100)}$
    \State $optimumK \gets -1$
    \State $minTime \gets \infty$
    
    \For{$k \gets 1$ \textbf{to} $n$}
        \State $totalTime \gets 0$
        \For{\textbf{each} $mcb$ \textbf{in} $test$}
            \State $startTime \gets \text{CurrentTime()}$
            \State $res \gets \text{\textbf{\textsc{BSMap++}}(beaconStart, beaconEnd, mcb, k)}$
            \State $endTime \gets \text{CurrentTime()}$
            \State $totalTime \gets totalTime + (endTime - startTime)$
        \EndFor
        \If{$totalTime < minTime$}
            \State $minTime \gets totalTime$
            \State $optimumK \gets k$
        \EndIf
    \EndFor
    
    \State \Return $optimumK$
\EndProcedure
\end{algorithmic}
\end{algorithm}

The \textsc{BSMap} algorithm determines the slot number of a specific beacon block associated with a main chain block (see Algorithm~\ref{alg:BSMap}). It initializes variables m and n, representing the lower and upper bounds of the search range, respectively, and mcb, representing the main chain block number \textbf{\textit{(Line 2-4)}}. The algorithm utilizes a binary search, repeatedly computing the midpoint (\textit{beaconSlot}) between the current lower and upper bounds \textbf{\textit{(Line 6)}}. It then fetches the block number corresponding to the midpoint using the \emph{fetchBlockNumber} function \textbf{\textit{(Line 7)}}. The search range is adjusted by comparing the main chain block number (\textit{mcb}) and the fetched block number from the beacon chain. If the two block numbers match, the algorithm returns the corresponding slot number (i.e., \emph{beaconSlot}) \textbf{\textit{(Line 9)}}. If mcb is greater, the lower bound is updated to \emph{beaconSlot+1}; otherwise, the upper bound is updated to \emph{beaconSlot–1} \textbf{\textit{(Line 10-13)}}. This process continues until the search range is exhausted (i.e., when m exceeds n).  If no slot corresponding to the main chain block number is found, the algorithm returns -1 \textbf{\textit{(Line 16)}}. Using a binary search strategy, the algorithm efficiently narrows down the slot number associated with the main chain block.

The \textsc{BSMap++} algorithm is an optimized variant of the \textsc{BSMap} algorithm, designed to efficiently identify the slot number of a specific beacon block that corresponds to a main chain block. This is achieved by partitioning the search range into k partitions, thus enhancing search efficiency. It initializes variables m and n, representing the lower and upper bounds of the search range, respectively, and mcb, representing the main chain block number \textbf{\textit{(Line 2-4)}}. 
The algorithm includes a special condition for $k \leq 1$, where it performs a linear search over the entire range from m to n, checking each slot sequentially until it finds a matching block number or exhausts the search range, returning -1 if no match is found \textbf{\textit{(Line 5-12)}}.
 Then the algorithm uses \textbf{\textit{k-partitioned search}}, repeatedly to calculate the size of each partition, referred to as \emph{binSize}, divides the search range into \emph{k} partitions, and the algorithm retrieves beacon chain block numbers, referred to as (\emph{bcbs}) for the endpoints of these partitions (denoted as beaconSlots) \textbf{\textit{(Line 14-17)}}. In each iteration, the algorithm compares the main chain block number (mcb) with the fetched beacon chain block numbers; if a match is found, the algorithm returns the corresponding slot number \textbf{\textit{(Line 19-21)}}. If the beacon chain block number exceeds the main chain block number, the search range is adjusted to focus on the previous partition, thereby narrowing down the slots to be checked \textbf{\textit{(Line 23-25)}}. If no match is found within the current partitions, the loop exits, and the algorithm returns -1 \textbf{\textit{(Line 30-34)}}. This \textbf{\textit{k-partitioned search}} strategy enhances efficiency by reducing the number of slots examined in each iteration.

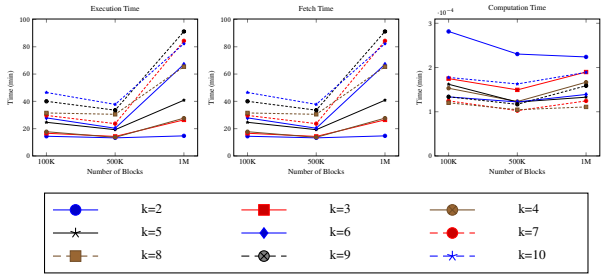
\begin{figure}
\centering
\begin{tikzpicture}[scale=0.32]
\begin{groupplot}[
    group style={
        group size=3 by 1, 
        horizontal sep=1.5cm, 
        vertical sep=1.5cm,
        group name=my plots
    },
    ymin=0,
    xtick=data,
    symbolic x coords={100K, 500K, 1M},
    xlabel={Number of Blocks},
    ylabel={Time (min)},
    legend to name=grouplegend,
    legend style={
        legend columns=3,
        anchor=north,
        /tikz/every even column/.append style={column sep=0.5cm},
        /tikz/every odd column/.append style={column sep=0.5cm},
        text width=0.75cm,
        font=\tiny,
    },
]

\nextgroupplot[
    title={Execution Time},
]

\addplot coordinates {(100K, 14.43626588) (500K, 13.26105892) (1M, 14.71963334)};
\addlegendentry{k=2}

\addplot coordinates {(100K, 16.63646101) (500K, 14.3033208) (1M, 26.42843665)};
\addlegendentry{k=3}

\addplot coordinates {(100K, 17.76972858) (500K, 13.64114064) (1M, 27.69811646)};
\addlegendentry{k=4}

\addplot coordinates {(100K, 24.65516181) (500K, 19.24664485) (1M, 40.83924683)};
\addlegendentry{k=5}

\addplot coordinates {(100K, 27.89264581) (500K, 20.30587318) (1M, 67.15619204)};
\addlegendentry{k=6}

\addplot coordinates {(100K, 29.70862357) (500K, 23.66881209) (1M, 84.32481744)};
\addlegendentry{k=7}

\addplot coordinates {(100K, 31.5085768) (500K, 30.51569284) (1M, 65.39840028)};
\addlegendentry{k=8}

\addplot coordinates {(100K, 40.04885327) (500K, 33.54316259) (1M, 91.21270658)};
\addlegendentry{k=9}

\addplot coordinates {(100K, 46.50436075) (500K, 37.77788825) (1M, 82.23170908)};
\addlegendentry{k=10}

\nextgroupplot[
    title={Fetch Time},
]

\addplot coordinates {(100K, 14.43598436) (500K, 13.26082845) (1M, 14.71940943)};
\addlegendentry{k=2}

\addplot coordinates {(100K, 16.63628607) (500K, 14.30317137) (1M, 26.42824688)};
\addlegendentry{k=3}

\addplot coordinates {(100K, 17.76957546) (500K, 13.64101763) (1M, 27.69794979)};
\addlegendentry{k=4}

\addplot coordinates {(100K, 24.6549997) (500K, 19.24652292) (1M, 40.83911396)};
\addlegendentry{k=5}

\addplot coordinates {(100K, 27.89251226) (500K, 20.30575102) (1M, 67.15605301)};
\addlegendentry{k=6}

\addplot coordinates {(100K, 29.7084989) (500K, 23.66870963) (1M, 84.32469296)};
\addlegendentry{k=7}

\addplot coordinates {(100K, 31.50845655) (500K, 30.51558827) (1M, 65.39828939)};
\addlegendentry{k=8}

\addplot coordinates {(100K, 40.0487195) (500K, 33.5430458) (1M, 91.21254784)};
\addlegendentry{k=9}

\addplot coordinates {(100K, 46.50418319) (500K, 37.77772544) (1M, 82.23152004)};
\addlegendentry{k=10}

\nextgroupplot[
    title={Computation Time},
]

\addplot coordinates {(100K, 0.000281524) (500K, 0.000230464) (1M, 0.000223914)};
\addlegendentry{k=2}

\addplot coordinates {(100K, 0.000174938) (500K, 0.000149432) (1M, 0.000189763)};
\addlegendentry{k=3}

\addplot coordinates {(100K, 0.000153127) (500K, 0.000123013) (1M, 0.000166674)};
\addlegendentry{k=4}

\addplot coordinates {(100K, 0.000162105) (500K, 0.000121926) (1M, 0.00013287)};
\addlegendentry{k=5}

\addplot coordinates {(100K, 0.000133551) (500K, 0.000122155) (1M, 0.00013903)};
\addlegendentry{k=6}

\addplot coordinates {(100K, 0.000124669) (500K, 0.000102452) (1M, 0.000124474)};
\addlegendentry{k=7}

\addplot coordinates {(100K, 0.000120249) (500K, 0.000104567) (1M, 0.000110896)};
\addlegendentry{k=8}

\addplot coordinates {(100K, 0.000133774) (500K, 0.000116793) (1M, 0.000158741)};
\addlegendentry{k=9}

\addplot coordinates {(100K, 0.000177556) (500K, 0.000162813) (1M, 0.000189037)};
\addlegendentry{k=10}

\end{groupplot}

\node at ($(my plots c1r1.south)!0.5!(my plots c3r1.south)-(0cm,1.5cm)$) [inner sep=0pt,anchor=north] {\ref{grouplegend}};

\end{tikzpicture}
\caption{\textsc{BSMap++} performance on varying workloads and number of partitions(k)}
\label{fig:BSMap++_Performance}
\end{figure}

The \textsc{ComputeOptimumK} algorithm determines the optimal number of partitions(k) for the \textsc{BSMap++} algorithm to achieve the best performance (see Algorithm~\ref{alg:findOptimumK}). It starts by generating a test set of 10 random block numbers within the main chain block range \textbf{\textit{(Line 2)}}. The algorithm initializes the optimumK and minimum time to default values \textbf{\textit{(Line 3-4)}}. For each possible k value from 1 to n, it measures the total time taken to execute \textsc{BSMap++} for all test values to calculate the execution time \textbf{\textit{(Line 7-12)}}. If the total execution time for the current k is less than the minimum recorded time, it updates the minimum time and the optimal k value \textbf{\textit{(Line 13-16)}}. Finally, it returns the optimal value of k \textbf{\textit{(Line 18)}}. This process ensures that the chosen kk minimizes the execution time of \textsc{BSMap++}, thereby optimizing performance. In our case, \emph{k=8}, \emph{k=7}, and \emph{k=8} are the optimal choices for 100K, 500K, and 1M blocks respectively (see Figure \ref{fig:BSMap++_Performance} for further details).

\begin{equation} \label{eqn:zscore}
        \emph{z-score} = \frac{x-\mu}{\sigma}
\end{equation}

To enhance data quality, we conducted a thorough analysis using statistical methods. We systematically removed rows with missing values to ensure a more comprehensive dataset. Subsequently, we addressed outliers, which were frequently encountered, by using \emph{z-score} (see Eq.~\ref{eqn:zscore} where x represents a data point, $\mu$ is the mean, and $\sigma$ is the standard deviation) and the Standardized Interquartile Range (\emph{$IQR_{standardized}$}) (see Eq.~\ref{eqn:IQR} where $Q_{25}$, $Q_{50}$ and $Q_{75}$ represent the $25^{th}$, $50^{th}$, and $75^{th}$ percentiles of attribute y in transactions).


\begin{equation} \label{eqn:IQR}
        IQR_{standardized} = \frac{Q_{75}(y) - Q_{25}(y)}{Q_{50}(y)}
\end{equation}

\emph{z-score} and \emph{standardized IQR} are chosen for their resilience to outliers compared to standard measures like mean and standard deviation. Utilizing these methods effectively identified and removed outliers from the dataset, resulting in a cleaner and more reliable dataset for further analysis.

\section{Empirical Results}\label{emp}
This section presents a performance study of \textsc{BSMap++} algorithm and empirical results answering two questions. First, how does \emph{EIP-3675} affect miner dynamics? (see Section~\ref{minerDA}). Second, how do \texttt{\textsl{Ram}} determine \emph{txnFee} (i.e., \emph{priorityFee} and \emph{baseFee}) and \emph{txnTime}? (see Section~\ref{ml}). For the experimental setup, we utilized a local Desktop system with a $12^{th}$ Gen Intel(R) Core (TM) i7-12700 processor running at 2.10 GHz, 32 GB RAM, 256 GB SSD, and one TB HDD. 



\begin{table}[]
\centering
\caption{Miner frequency comparison}
\label{tab:freq_comp}
\begin{tabular}{ccl}
\hline
\multicolumn{1}{|c|}{\multirow{2}{*}{\textbf{Number of Blocks}}} & \multicolumn{2}{c|}{\textbf{Number of Miners/Proposers}}                                \\ \cline{2-3} 
\multicolumn{1}{|c|}{}                                           & \multicolumn{1}{c|}{\textbf{PoW-Ethereum}} & \multicolumn{1}{c|}{\textbf{PoS-Ethereum}} \\ \hline
\multicolumn{1}{|c|}{\textbf{100K}}                              & \multicolumn{1}{c|}{72 (1x)}               & \multicolumn{1}{c|}{2689 (37.3x)}          \\ \hline
\multicolumn{1}{|c|}{\textbf{500K}}                              & \multicolumn{1}{c|}{91 (1x)}               & \multicolumn{1}{c|}{5596 (61.5x)}          \\ \hline
\multicolumn{1}{|c|}{\textbf{1M}}                                & \multicolumn{1}{c|}{116 (1x)}              & \multicolumn{1}{c|}{7115 (61.3x)}          \\ \hline
\end{tabular}
\end{table}

\subsection{\textsc{BSMap++} Dynamics}\label{BSMapDA}
In our analysis, we utilized block data ranging from block number 18000001 to 19000000, segmented into three distinct volumes: 100K, 500K, and 1M blocks. Concurrently, we selected beacon slots spanning from slot number 7100000 to 8200000. To identify the optimal value of k (i.e., optimal number of partitions) for our workloads, we implemented the \emph{\textsc{ComputeOptimumK}} algorithm (see Algorithm \ref{alg:findOptimumK}), which facilitated the computation of three primary metrics: total execution time, total fetch time, and total computation time. The computation time is derived by subtracting the total fetch time from the total execution time. It is important to note that the total fetch time fluctuates due to the variability in each request made to the local host for retrieving block numbers corresponding to the beacon slot numbers. We visualized the performance results through three separate line charts—representing execution time, fetch time, and computation time—by varying k from 2 to 10 for all three dataset volumes. The results indicated that for the 100K block dataset, the optimal k value was 8, resulting in the minimum computation time. For the 500K block dataset, the optimal k was determined to be 7, whereas for the 1M block dataset, the optimal k reverted to 8, indicating the least computation time for these specific volumes (see Figure \ref{fig:BSMap++_Performance}).

\subsection{Miner Dynamics}\label{minerDA}
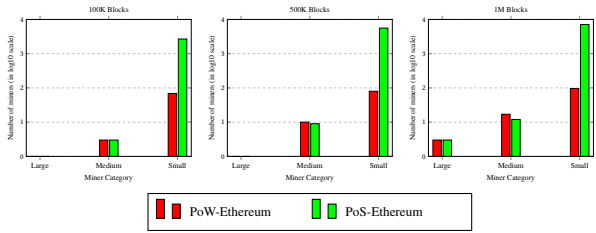
\begin{figure}
\centering
\begin{tikzpicture}[scale=0.32]
\begin{groupplot}[
    group style={
        group size=3 by 1, 
        horizontal sep=1.5cm, 
        vertical sep=1.5cm,
        group name=my plots
    },
    ybar,
    ymin=0, 
    ymax=4,
    ymajorgrids=true,
    grid style=dashed,
    symbolic x coords={Large, Medium, Small}, 
    xtick=data, 
    legend to name=grouplegend, 
    legend style={ 
        legend columns=-1,
        anchor=north, 
        /tikz/every even column/.append style={column sep=0.05cm},
        /tikz/every odd column/.append style={column sep=0.05cm},
        text width=1.5cm, 
        font=\tiny, 
    },
]
\nextgroupplot[
    title={100K Blocks},
    xlabel={Miner Category},
    ylabel={Number of miners (in log10 scale)},
]
\addplot[fill=red] coordinates {(Large, 0) (Medium, 0.477121255) (Small, 1.838849091)};
\addplot[fill=green] coordinates {(Large, 0) (Medium, 0.477121255) (Small, 3.429106008)};

\addlegendentry{PoW-Ethereum}
\addlegendentry{PoS-Ethereum}

\nextgroupplot[
    title={500K Blocks},
    xlabel={Miner Category},
    ylabel={Number of miners (in log10 scale)},
]
\addplot[fill=red] coordinates {(Large, 0) (Medium, 1) (Small, 1.903089987)};
\addplot[fill=green] coordinates {(Large, 0) (Medium, 0.954242509) (Small,3.747100931)};

\nextgroupplot[
    title={1M Blocks},
    xlabel={Miner Category},
    ylabel={Number of miners (in log10 scale)},
]
\addplot[fill=red] coordinates {(Large, 0.477121255) (Medium, 1.230448921) (Small, 1.982271233)};
\addplot[fill=green] coordinates {(Large, 0.477121255) (Medium, 1.079181246) (Small, 3.851258349)};

\addlegendentry{PoW-Ethereum}
\addlegendentry{PoS-Ethereum}

\end{groupplot}

\node at ($(my plots c1r1.south)!1!(my plots c2r1.south)-(0cm,1.5cm)$) [inner sep=0pt,anchor=north] {\ref{grouplegend}};

\end{tikzpicture}
\caption{Miner Categorization}
\label{fig:categorization}
\end{figure}

\begin{figure}
\centering
\begin{tikzpicture}[scale=0.4]
\begin{groupplot}[
    group style={
        group size=2 by 1,
        horizontal sep=3cm,
    },
    width=0.5\textwidth,
    height=0.4\textwidth,
    ymin=0,
    ymax=11,
    ytick=\empty, 
    xtick=\empty,
    xlabel={Block number},
    xmajorgrids,
    enlarge x limits=false,
    grid style=dashed,
    scaled x ticks=false, 
    ]

\nextgroupplot[
    title={PoW-Ethereum},
    ytick={1,2,3,4,5,6,7,8,9,10},
    yticklabels={\texttt{0xea6..},\texttt{0x829..},\texttt{0x1ad..},\texttt{0x001..},\texttt{0x7f1..},\texttt{0x2da..},\texttt{0x52b..},\texttt{0xc73..},\texttt{0x646..},\texttt{0x3ec..}}, 
    yticklabel style={text width=4em,align=left},
    ylabel={Miner's public key},
    xtick={15534900, 15535734, 15536467, 15537400},
    xticklabels={15534900, 15535734, 15536467, 15537400},
    xticklabel style={
        /pgf/number format/fixed,
        /pgf/number format/fixed zerofill,
        /pgf/number format/precision=5,
    },
    enlarge x limits={value=0.05, auto=true},
    enlarge y limits={value=0.05, auto=true},
]

\pgfplotstableread[col sep=comma]{data/PoW-Eth.csv}\loadedtable
\foreach \i in {1,...,10} {
    \addplot[scatter, only marks] table [x=x\i, y=y\i, col sep=comma] {\loadedtable};
}
\nextgroupplot[
    title={PoS-Ethereum},
    ytick={1,2,3,4,5,6,7,8,9,10},
    yticklabels={\texttt{0xdaf..},\texttt{0x690..},\texttt{0x952..},\texttt{0x388..},\texttt{0x467..},\texttt{0xf2f..},\texttt{0xfee..},\texttt{0x473..},\texttt{0x199..},\texttt{0xaab..}}, 
    yticklabel style={text width=4em,align=left},
    ylabel={Proposer's public key},
    xtick={15537300, 15603667, 15710034 , 15796400},
    xticklabels={15537300, 15603667, 15710034 , 15796400},
    xticklabel style={
        /pgf/number format/fixed,
        /pgf/number format/fixed zerofill,
        /pgf/number format/precision=5,
    },
    enlarge x limits={value=0.05, auto=true}, 
    enlarge y limits={value=0.05, auto=true}, 
]

\pgfplotstableread[col sep=comma]{data/PoS-Eth.csv}\loadedtable
\foreach \i in {1,...,10} {
    \nextgroupplot
    \addplot[scatter, only marks] table [x=x\i, y=y\i, col sep=comma] {\loadedtable};
}

\end{groupplot}
\end{tikzpicture}
\caption{Ethereum Block Distribution}
\label{fig:Eth-block}
\end{figure}
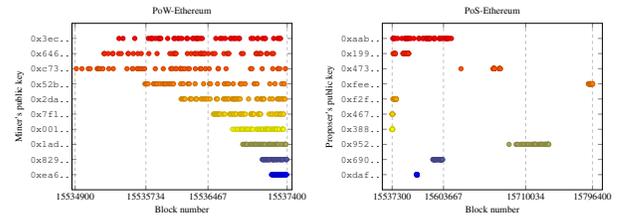

\textbf{\textit{High-Level Idea.}} \emph{EIP-3675} replaces computational resources with stakes, altering miners' dynamics. For fair assessment, we considered the last 1 M blocks of PoW-Ethereum and the first 1M blocks of PoS-Ethereum.  To begin with, we ran an experiment on the miner count. Results have indicated that the miner count has significantly increased for PoS-Ethereum, compared to PoW-Ethereum (see Table~\ref{tab:freq_comp}). To further understand the variation, we categorized miners/proposers into three categories, i.e., \emph{Small}, \emph{Medium}, and \emph{Large}, based on the number of blocks they have mined. The objective was to gain insights into the distribution of mining activities among peers in the network based on their resource-holding capabilities (computational resources in PoW and Stakes in PoS). The results have shown that the small-scale miner count has particularly seen a huge increase with little to no change in the rest of the two categories, i.e., \emph{Medium} and \emph{Large} (see Fig~\ref{fig:categorization}). Finally, we have explored the randomness of miner selection by analyzing the patterns of the last 50 blocks mined by top-10 miners for PoW-Ethereum and the first 50 blocks mined by top-10 proposers for PoS-Ethereum. The results show reduced randomness in miner selection, i.e., \emph{Large} proposers were able to mine blocks in a close-to-contiguous manner (see Fig~\ref{fig:Eth-block}).



\textbf{\textit{\underline{Miner Count Experiment.}}} We extracted and analyzed the unique number of miners/proposers from \emph{PoW-Ethereum} and \emph{PoS-Ethereum} networks across various block volumes (100K, 500K, and 1M blocks), respectively. The results unveiled a notable difference (see Table~\ref{tab:freq_comp}); \emph{PoS-Ethereum} network exhibited a significant increase ($\approx50\times$) in unique miners. It means \emph{PoS-Ethereum} encourages broader participation, resulting in a larger pool of unique proposers in \emph{PoS-Ethereum}.


\begin{table*}[h]
\footnotesize
\centering
\caption{Model performance comparison for \emph{baseFee} prediction}
\label{tab:baseFee}
\begin{tabular}{|c|lll|lll|lll|}
\hline
\textbf{Total} & \multicolumn{3}{c|}{\textbf{100K Transactions}}                                                                                                       & \multicolumn{3}{c|}{\textbf{200K Transactions}}                                                                                                       & \multicolumn{3}{c|}{\textbf{500K Transactions}}                                                                                                       \\ \hline
\textbf{Model} & \multicolumn{1}{c|}{\textbf{MAE}}                     & \multicolumn{1}{c|}{\textbf{RMSE}}                    & \multicolumn{1}{c|}{\textbf{R2Score}} & \multicolumn{1}{c|}{\textbf{MAE}}                     & \multicolumn{1}{c|}{\textbf{RMSE}}                    & \multicolumn{1}{c|}{\textbf{R2Score}} & \multicolumn{1}{c|}{\textbf{MAE}}                     & \multicolumn{1}{c|}{\textbf{RMSE}}                    & \multicolumn{1}{c|}{\textbf{R2Score}} \\ \hline
\textbf{ET}    & \multicolumn{1}{l|}{\cellcolor[HTML]{90EE90}13418.66} & \multicolumn{1}{l|}{39786.01}                         & 0.999002                              & \multicolumn{1}{l|}{\cellcolor[HTML]{90EE90}4854.707} & \multicolumn{1}{l|}{\cellcolor[HTML]{90EE90}14342.27} & \cellcolor[HTML]{90EE90}0.999843      & \multicolumn{1}{l|}{15728.64}                         & \multicolumn{1}{l|}{39284.53}                         & 0.99936                               \\ \hline
\textbf{KN}    & \multicolumn{1}{l|}{\cellcolor[HTML]{FE996B}182125}   & \multicolumn{1}{l|}{\cellcolor[HTML]{FE996B}320350.3} & \cellcolor[HTML]{FE996B}0.935301      & \multicolumn{1}{l|}{\cellcolor[HTML]{FE996B}126884}   & \multicolumn{1}{l|}{\cellcolor[HTML]{FE996B}234678.3} & \cellcolor[HTML]{FE996B}0.958051      & \multicolumn{1}{l|}{\cellcolor[HTML]{FE996B}388239.2} & \multicolumn{1}{l|}{\cellcolor[HTML]{FE996B}601938.4} & \cellcolor[HTML]{FE996B}0.849626      \\ \hline
\textbf{LR}    & \multicolumn{1}{l|}{126529.2}                         & \multicolumn{1}{l|}{177014.7}                         & 0.980246                              & \multicolumn{1}{l|}{122060.1}                         & \multicolumn{1}{l|}{169425.5}                         & 0.978136                              & \multicolumn{1}{l|}{358227.1}                         & \multicolumn{1}{l|}{480535.5}                         & 0.904166                              \\ \hline
\textbf{RF}    & \multicolumn{1}{l|}{14516.22}                         & \multicolumn{1}{l|}{40374.6}                          & 0.998972                              & \multicolumn{1}{l|}{6817.447}                         & \multicolumn{1}{l|}{17456.16}                         & 0.999768                              & \multicolumn{1}{l|}{\cellcolor[HTML]{90EE90}13039.19} & \multicolumn{1}{l|}{\cellcolor[HTML]{90EE90}30149.31} & \cellcolor[HTML]{90EE90}0.999623      \\ \hline
\textbf{GB}    & \multicolumn{1}{l|}{35695.44}                         & \multicolumn{1}{l|}{62800.32}                         & 0.997514                              & \multicolumn{1}{l|}{24969.59}                         & \multicolumn{1}{l|}{38985.51}                         & 0.998842                              & \multicolumn{1}{l|}{51570.78}                         & \multicolumn{1}{l|}{79860.5}                          & 0.997353                              \\ \hline
\textbf{XGB}   & \multicolumn{1}{l|}{15997.23}                         & \multicolumn{1}{l|}{\cellcolor[HTML]{90EE90}35573.88} & \cellcolor[HTML]{90EE90}0.999202      & \multicolumn{1}{l|}{10935.79}                         & \multicolumn{1}{l|}{20148.54}                         & 0.999691                              & \multicolumn{1}{l|}{27519.35}                         & \multicolumn{1}{l|}{50995.66}                         & 0.998921                              \\ \hline
\end{tabular}

\begin{minipage}{14.5cm}
\vspace{0.1cm}
\tiny 
Notes: (1) Green highlights indicate the lowest MAE, lowest RMSE, and highest R2 score (\emph{best results}). (2) Red highlights indicate the highest MAE, highest RMSE, and lowest R2 score (\emph{worst results}).
\end{minipage}

\end{table*}

\textbf{\textit{\underline{Miner Distribution Experiment.}}}
 We categorized miners/proposers into three categories: Large (miners who have mined more than 100K blocks), Medium (miners who have mined 10K to 100K blocks), and Small (miners who have mined less than 10K blocks). Following this categorization, we represented the distribution of miners across these categories (see Fig.~\ref{fig:categorization}). To improve the visualization's clarity and effectiveness, we applied a logarithmic scale to the `number of miners', recognizing the significant disparity between PoW and PoS data. The impact of \emph{EIP-3675} is apparent through the logarithmic representation, notably revealing a considerable increase ($\approx60\times$) in the number of \textit{\textbf{small-scale}} miners across block volumes (100K, 500K, and 1M). This observation highlights a substantial shift in miner dynamics in \emph{PoS-Ethereum}, emphasizing the heightened involvement of \textit{\textbf{small-scale}} miners. Our findings suggest that the transition to PoS creates a more flexible and accessible environment for miners, encouraging a diverse and decentralized network participation landscape.  \\

 
\textbf{\textit{\underline{Miner Randomness Experiment.}}} We designed the experiment for miner randomness as follows. First, we find the top 10 miners for PoW-Ethereum and the top 10 proposers for PoS-Ethereum. Later, we extract block numbers of the last 50 blocks mined by each of the top 10 miners for PoW-Ethereum. Similarly, we extract the block numbers of the first 50 blocks mined by each of the top 10 proposers for PoS-Ethereum. We generate two scatter plots to study miner randomness (see Fig.~\ref{fig:Eth-block}). The left plot depicts the last 50 blocks mined by each of the top 10 miners of \emph{PoW-Ethereum}, while the right plot depicts the first 50 blocks proposed by each of the top 10 proposers of \emph{PoS-Ethereum}. For improved clarity, we have truncated the 256-bit public key to display only the first five characters. Clearly, the block proposals by proposers in \emph{PoS-Ethereum} exhibit a higher degree of consecutiveness or alternativeness than \emph{PoW-Ethereum}. This observation shows the reduction in block proposals’ randomness among block proposers, as PoW relies on computational puzzle-solving, which is more random than staking. Despite the multifold increase in the number of \textit{\textbf{small-scale}} miners, there is a simultaneous decrease in the overall randomness of the system.

\subsection{Transaction Fee and Time Prediction}\label{ml}
The \emph{txnFee} in Ethereum comprises two crucial components: the \emph{baseFee} and the \emph{priorityFee}. Determining the \emph{baseFee} considers various factors such as network congestion, transaction confirmation demand, and the availability of active validators. On the other hand, users can choose the \emph{priorityFee}. While it is technically possible to send a transaction without a \emph{priorityFee}, in practical terms, proposers are unlikely to accept such transactions. The \emph{baseFee} is burnt during transaction processing, while the \emph{priorityFee} goes to the proposer. Deciding \emph{priorityFee} becomes important as users must maintain a balance i.e., paying too little may result in transaction delays or non-inclusion, while spending too much may lead to unnecessary expenses, as the \emph{priorityFee} can only offer benefits up to a certain extent. 

To address this challenge, we determine an optimal \emph{txnFee} that ensures a proposer accepts a user's transaction. Our approach aims to aid the fee-setting process by utilizing regression-based machine learning models. We train a set of machine learning regression models on a diverse set of data that encompasses network conditions, validator activities, and historical transaction patterns to recognize complicated patterns and correlations between them. The machine learning models serve as valuable tools for users, providing insights into the optimal combination of \emph{baseFee} and \emph{priorityFee} for successful and timely transaction confirmations.

\textbf{\textit{High-Level Idea.}} We have employed the regression prediction models as in~\cite{DBLP:conf/dsa/LiuWLXG19, DBLP:journals/sigmetrics/OliveiraVSVBVG21, DBLP:conf/compsac/Mars0CK21, DBLP:journals/comsur/LiuYLJL20, DBLP:conf/bigdataconf/YinV17}, to choose the most suitable model for predicting \emph{txnFee} and \emph{txnTime} accurately for PoS-Ethereum. These models include Extra Trees Regressor (\emph{ET}), K-Neighbours Regressor (\emph{KN}), Linear Regression (\emph{LR}), Random Forest Regressor (\emph{RF}), Gradient Boosting Regressor (\emph{GB}), and Extreme  Gradient Boosting (\emph{XGB}). Each model brought unique strengths and perspectives, allowing for the exploration of the complex TFM dynamics. To compare these models, we have also designed a baseline \emph{txnFee} estimation method.

\textbf{\textit{\underline{Baseline Transaction Fee Estimation Method.}}} The state-of-the-art mostly discusses machine learning models trained on transaction-related data (based on PoW) for predicting of \emph{txnFee} in PoW-Ethereum setting. On the contrary, we fed ML models with network-related data (PoS-based) and transaction-related data for making predictions in PoS-Ethereum setting. Therefore, a direct comparison is not feasible. However, as it is a critical aspect, we have included the comparison against the Etherscan's estimation method named `Ethereum Gas Tracker'. \emph{Etherscan} (a Block Explorer and Analytics Platform for Ethereum)~\cite{EtherScan}, has an inbuilt estimation approach to predict the transaction confirmation time based on the gas price setted by the user. The estimation approach in \emph{Etherscan} utilizes the gas price to group transactions of the last 1000 blocks and calculates the average confirmation time. Simlar to this estimation approach, we have established a baseline by calculating the mean \emph{txnFee} by taking the avearage of the last 1000 transaction fees (see Eq.~\ref{eqn:estimate}). This has provided a starting point for comparison and set the stage for a more sophisticated ML approaches to predict \emph{txnFee}. We trained a set of regression-based machine learning models, and results are compared against the baseline approach.

\begin{equation} \label{eqn:estimate}
txnFee_n = \frac{1}{1000} \sum_{i=n-1000}^{n-1} txnFee_i
\end{equation}

We have considered the following evaluation metrics to assess the performance of the regression-based ML models: Root Mean Squared Error (RMSE), Mean Absolute Error (MAE), and R-squared ($R^2$). These metrics provided a quantitative measure of accuracy, enabling us to discern the strengths and weaknesses of each model in capturing the underlying patterns in \emph{txnfee} and \emph{txnTime}.

\textbf{RMSE} measures the deviation between actual and predicted values and accurately reflects the measurement's precision. The prediction's accuracy is inversely proportional to the \emph{RMSE} value; a lower \emph{RMSE} value indicates higher accuracy.

\begin{equation} \label{eqn:rmse}
RMSE = \sqrt{\frac{1}{n} \sum_{i=i}^{n} (y_i - \hat{y})^2}
\end{equation}

It is calculated as the root of the mean value of the squared difference between actual value and predicted value, as shown in Eq.~\ref{eqn:rmse}. Here, $y_i$ and $\hat{y}_i$ are the actual and predicted values for the $i^{th}$ transaction, and $n$ is the number of predicted values.

\textbf{MAE} is calculated by taking the mean of the absolute difference between the actual and predicted values, as shown in Eq. ~\ref{eqn:mae}. Similar to \emph{RMSE}, the lower \emph{MAE} value reflects the higher prediction accuracy.

\begin{equation} \label{eqn:mae}
MAE = \frac{1}{n} \sum_{i=i}^{n} \mid y_i - \hat{y}\mid
\end{equation}

\boldmath $R^2$ \unboldmath \textbf{Score (coefficient of determination)} quantifies the extent of variation in the dependent output value predicted using the independent input value. Essentially, it serves as a metric to gauge the effectiveness of a regression model in forecasting actual data outcomes. The \emph{$R^2$ Score} is proportional to the model's accuracy; a higher \emph{$R^2$ Score} is preferable as it signifies better results.

\begin{equation} \label{eqn:r2}
R^2 = 1 - \frac{SS_{res}}{SS_{tot}} = 1 - \frac{\sum_{i=1}^{n}{(y_i - \hat{y_i})^2}}{\sum_{i=1}^{n}{(y_i - \Bar{y_i})^2}}
\end{equation}

It is calculated as one minus the ratio of the sum of squared residuals to the total sum of squares, as shown in Eq.~\ref{eqn:r2}. \emph{$SS_{res}$} is the sum of squared residuals, \emph{$SS_{tot}$} is the total sum of squares, and $\Bar{y}$ is the mean of the actual values.

For \emph{baseFee} prediction, \emph{ET} and \emph{RF} perform well for \emph{200K} and \emph{500K} transactions respectively (see Table~\ref{tab:baseFee}). Both exhibit low \emph{MAE} and \emph{RMSE} values and a high \emph{$R^2$} value, indicating high accuracy in predicting the \emph{baseFee}. However, \emph{KN} displays comparatively lower accuracy metrics across all transaction volumes, yielding higher\emph{MAE} and \emph{RMSE} values and a lower \emph{$R^2$} value, indicating that it might not be as well-suited for fee prediction as compared to other models.



\begin{figure}[h]
 \centering
  \subfloat[Best]{\includegraphics[width=.45\linewidth]{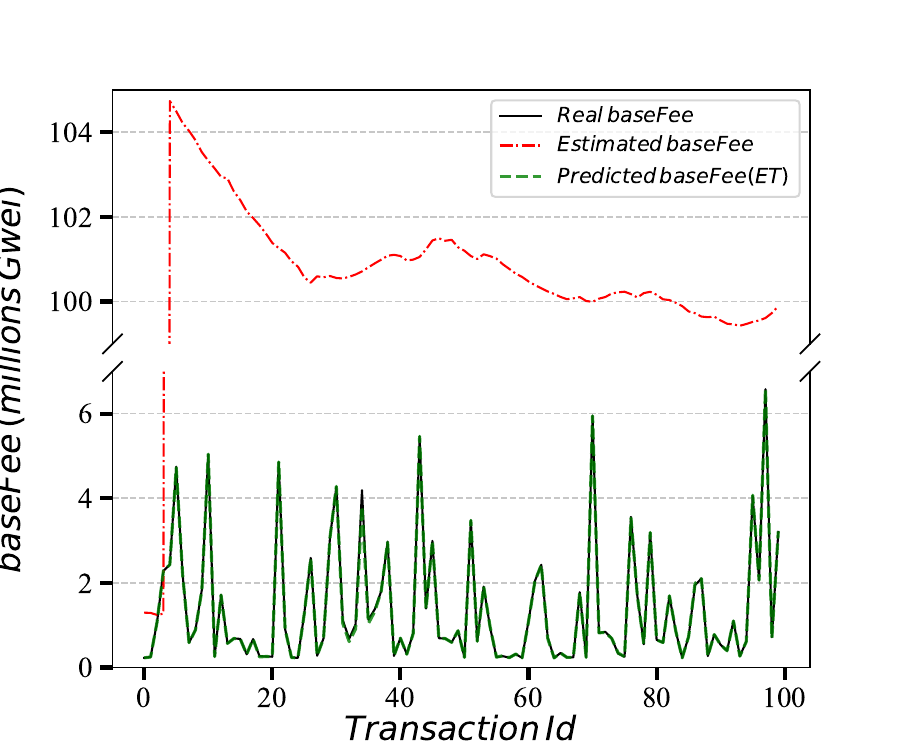}}
  \subfloat[Worst]{\includegraphics[width=.45\linewidth]{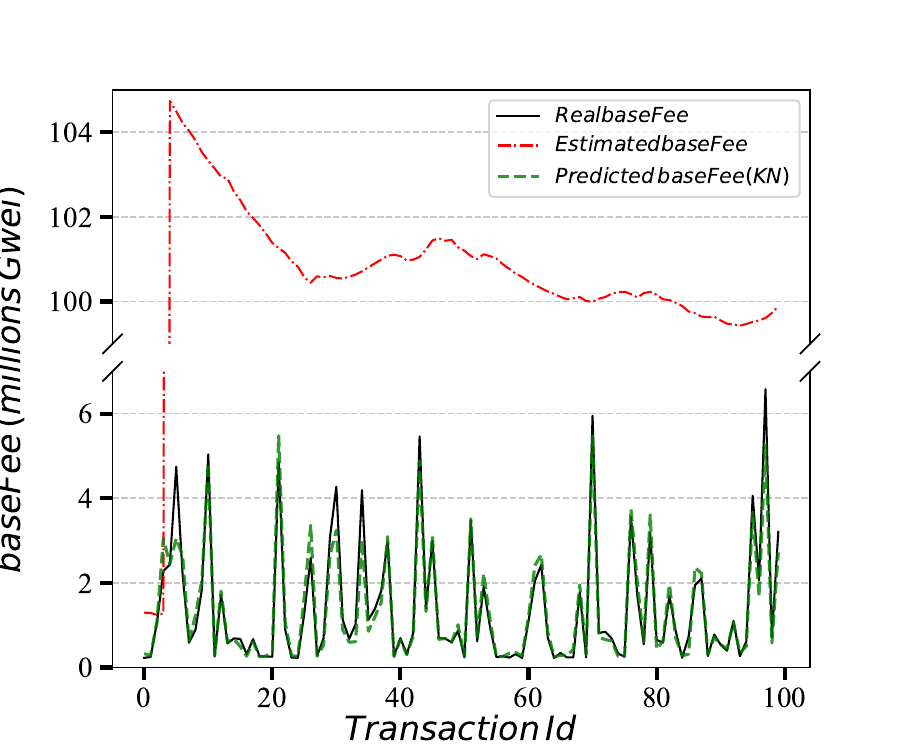}}
  \caption{Prediction results for \emph{baseFee}}
  \label{fig:baseFee}
\end{figure}

\begin{table*}[h]
\footnotesize
\centering
\caption{Model performance comparison for \emph{priorityFee} prediction}
\label{tab:priorityFee}
\begin{tabular}{|c|lll|lll|lll|}
\hline
\textbf{}      & \multicolumn{3}{c|}{\textbf{100K Transactions}}                                                                                                       & \multicolumn{3}{c|}{\textbf{200K Transactions}}                                                                                                       & \multicolumn{3}{c|}{\textbf{500K Transactions}}                                                                                                       \\ \hline
\textbf{Model} & \multicolumn{1}{c|}{\textbf{MAE}}                     & \multicolumn{1}{c|}{\textbf{RMSE}}                    & \multicolumn{1}{c|}{\textbf{R2Score}} & \multicolumn{1}{c|}{\textbf{MAE}}                     & \multicolumn{1}{c|}{\textbf{RMSE}}                    & \multicolumn{1}{c|}{\textbf{R2Score}} & \multicolumn{1}{c|}{\textbf{MAE}}                     & \multicolumn{1}{c|}{\textbf{RMSE}}                    & \multicolumn{1}{c|}{\textbf{R2Score}} \\ \hline
\textbf{ET}    & \multicolumn{1}{l|}{\cellcolor[HTML]{90EE90}5719.28}  & \multicolumn{1}{l|}{17311.48}                         & 0.948008                              & \multicolumn{1}{l|}{\cellcolor[HTML]{90EE90}5016.343} & \multicolumn{1}{l|}{17523.39}                         & 0.939837                              & \multicolumn{1}{l|}{10298.51}                         & \multicolumn{1}{l|}{24332.18}                         & 0.88461                               \\ \hline
\textbf{KN}    & \multicolumn{1}{l|}{\cellcolor[HTML]{FE996B}48590.26} & \multicolumn{1}{l|}{\cellcolor[HTML]{FE996B}78209.76} & \cellcolor[HTML]{FE996B}-0.06119      & \multicolumn{1}{l|}{\cellcolor[HTML]{FE996B}43888.42} & \multicolumn{1}{l|}{\cellcolor[HTML]{FE996B}72811.24} & \cellcolor[HTML]{FE996B}-0.03869      & \multicolumn{1}{l|}{\cellcolor[HTML]{FE996B}52302.45} & \multicolumn{1}{l|}{\cellcolor[HTML]{FE996B}80058.58} & \cellcolor[HTML]{FE996B}-0.24917      \\ \hline
\textbf{LR}    & \multicolumn{1}{l|}{29119.9}                          & \multicolumn{1}{l|}{48814.51}                         & 0.5866                                & \multicolumn{1}{l|}{26587.31}                         & \multicolumn{1}{l|}{46005.95}                         & 0.585314                              & \multicolumn{1}{l|}{29095.41}                         & \multicolumn{1}{l|}{45330.49}                         & 0.599515                              \\ \hline
\textbf{RF}    & \multicolumn{1}{l|}{5803.747}                         & \multicolumn{1}{l|}{17319.83}                         & 0.947957                              & \multicolumn{1}{l|}{5540.436}                         & \multicolumn{1}{l|}{19328.37}                         & 0.926805                              & \multicolumn{1}{l|}{\cellcolor[HTML]{90EE90}10178.88} & \multicolumn{1}{l|}{24094.48}                         & 0.886854                              \\ \hline
\textbf{GB}    & \multicolumn{1}{l|}{6507.821}                         & \multicolumn{1}{l|}{\cellcolor[HTML]{90EE90}16897.28} & \cellcolor[HTML]{90EE90}0.950466      & \multicolumn{1}{l|}{6187.508}                         & \multicolumn{1}{l|}{\cellcolor[HTML]{90EE90}16742.03} & \cellcolor[HTML]{90EE90}0.945083      & \multicolumn{1}{l|}{10275.47}                         & \multicolumn{1}{l|}{\cellcolor[HTML]{90EE90}23707.63} & \cellcolor[HTML]{90EE90}0.890458      \\ \hline
\textbf{XGB}   & \multicolumn{1}{l|}{6866.978}                         & \multicolumn{1}{l|}{19180.89}                         & 0.936172                              & \multicolumn{1}{l|}{6186.524}                         & \multicolumn{1}{l|}{18394.71}                         & 0.933706                              & \multicolumn{1}{l|}{10980.68}                         & \multicolumn{1}{l|}{25443.89}                         & 0.873825                              \\ \hline
\end{tabular}
\begin{minipage}{14.5cm}
\vspace{0.1cm}
\tiny 
Notes: (1) Green highlights indicate the lowest MAE, lowest RMSE, and highest R2 score (\emph{best results}). (2) Red highlights indicate the highest MAE, highest RMSE, and lowest R2 score (\emph{worst results}).
\end{minipage}
\end{table*}

\textbf{\textit{\underline{Base Fee Prediction.}}} Complementing the quantitative metrics presented in Table~\ref{tab:baseFee}, line charts provide a dynamic visualization of the performance of the EtherScan estimation method and the machine learning models. We have plotted line charts for the best and worst models regarding performance, respectively, illustrating the relationship between actual, estimated, and predicted values for \emph{baseFee} (see Fig. \ref{fig:baseFee}). The x-axis represents transaction IDs, while the y-axis signifies the corresponding \emph{baseFee} values (in millions). We have defined a broken y-axis between 7M and 99M to provide clear visuals even for outliers (in the case of estimated \emph{baseFee}). Both charts focus on a systematic sample of 100 transactions for clear visualization. The plot displays three lines— the first line indicating actual \emph{baseFee} values (in black), the second line indicating estimated \emph{baseFee} values (in red dashes), and the third line depicting predicted values (in green dashes) from the machine learning models. The alignment among these lines highlights the model's accuracy (less deviation indicates high accuracy). Noticeably, the \emph{ET} performed well with negligible deviation (see Fig. \ref{fig:baseFee}(a)), while the \emph{KN} exhibited little deviation(see Fig. \ref{fig:baseFee}(b)). In contrast, the estimation method shows too much deviation from the actual values. Our proposed models, even the worst-performing model, yield highly accurate results compared to the estimation method.

\begin{figure}[h]
 \centering
  \subfloat[Best]{\includegraphics[width=.45\linewidth]{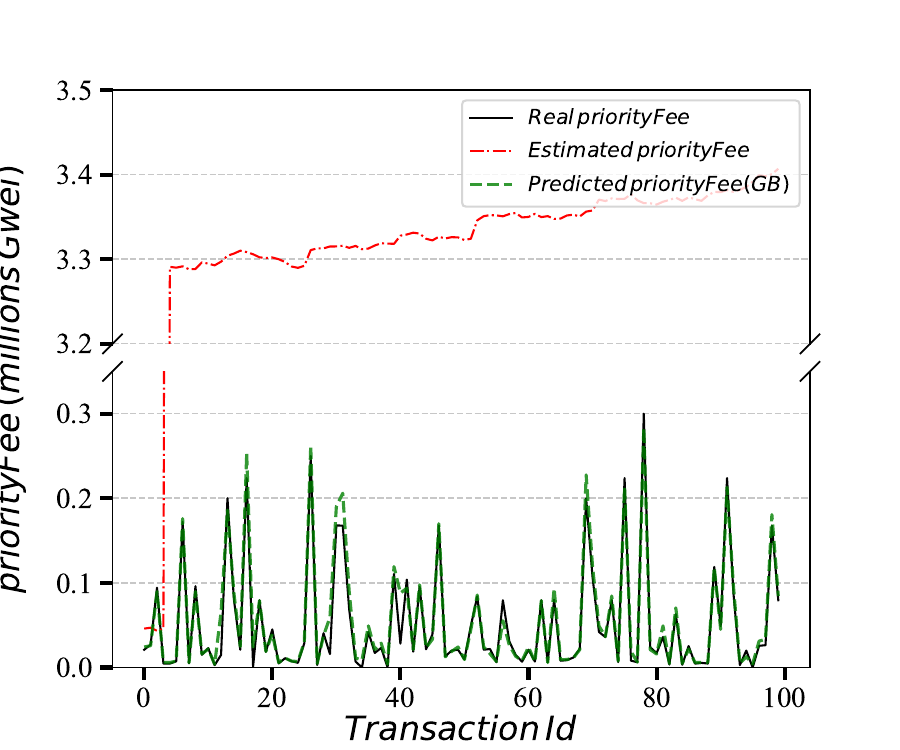}}
  \subfloat[Worst]{\includegraphics[width=.45\linewidth]{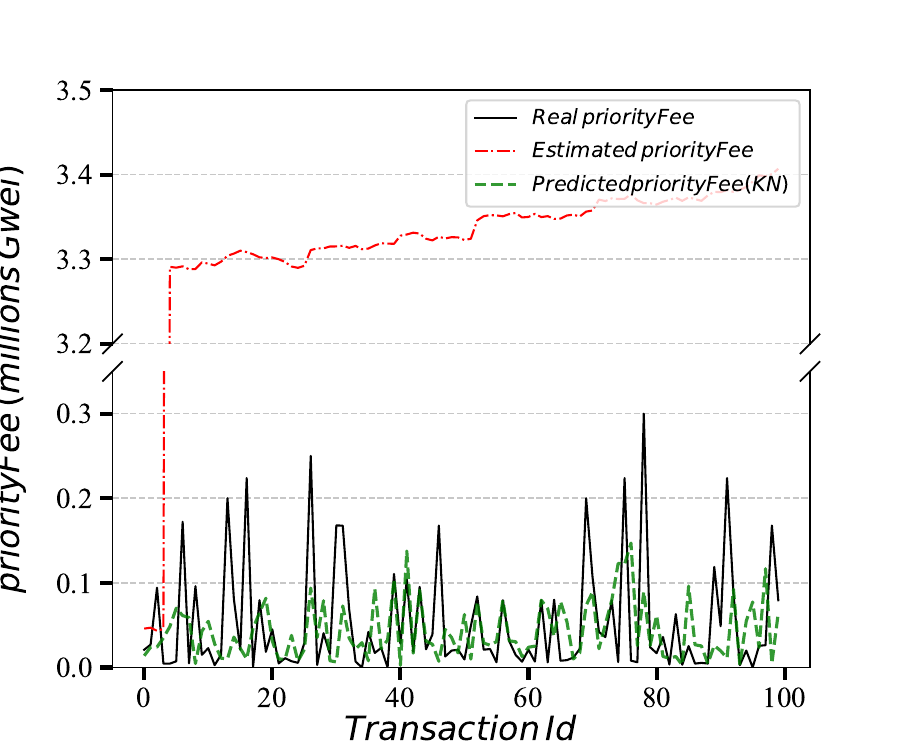}}
  \caption{Prediction results for \emph{priorityFee}}
  \label{fig:priorityFee}
\end{figure}

\textbf{\textit{\underline{Priority Fee Prediction.}}} Regarding \emph{priorityFee}, model efficiency is slightly different; no model performs exceptionally well across all transaction volumes (see Table~\ref{tab:priorityFee}). However, the \emph{GB} performed well in most metrics, with lower \emph{RMSE} and \emph{MAE} values and a higher \emph{$R^2$} value, indicating high accuracy in prediction results. Conversely, the \emph{KN} performs worst, similar to the case of the \emph{baseFee}. Similar to \emph{baseFee}, we have plotted the line charts for the best and worst models in terms of performance, respectively, illustrating the relationship between real, estimated, and predicted values for \emph{priorityFee} (see Fig.~\ref{fig:priorityFee}). 

The x-axis represents transaction IDs, while the y-axis signifies the corresponding \emph{priorityFee} values (in Thousands). We have used a broken y-axis between 350K and 3.1M to provide clear visuals even for outliers (in case of estimated \emph{priorityFee}). Similar to \emph{baseFee} prediction, we focus on a systematic sample of 100 transactions for clarity. Each chart showcases three lines— the first line indicating actual \emph{priorityFee} values (in black), the second line indicating estimated \emph{priorityFee} values (in red dashes), and the third line depicting predicted values (in green dashes) from the machine learning models. The coherence between these lines highlights their precision in predicting \emph{priorityFee}. Notably, the \emph{GB} shows little deviation (see Fig.~\ref{fig:priorityFee}(a)), and the \emph{KN} exhibits substantial deviations (see Fig.~\ref{fig:priorityFee}(b)), whereas the estimated values show extreme deviation from actual values.

\begin{figure}[h]
 \centering
  \subfloat[Best]{\includegraphics[width=.45\linewidth]{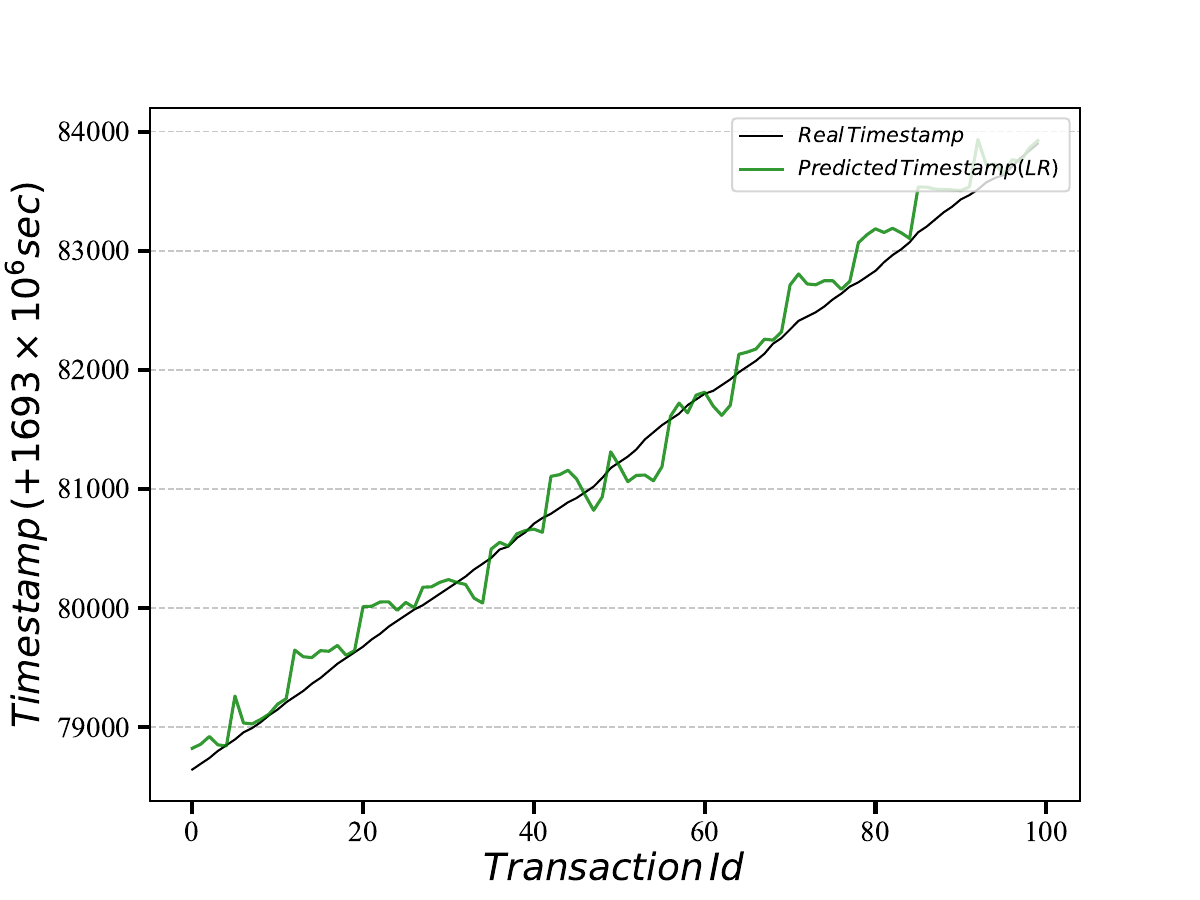}}
  \subfloat[Worst]{\includegraphics[width=.45\linewidth]{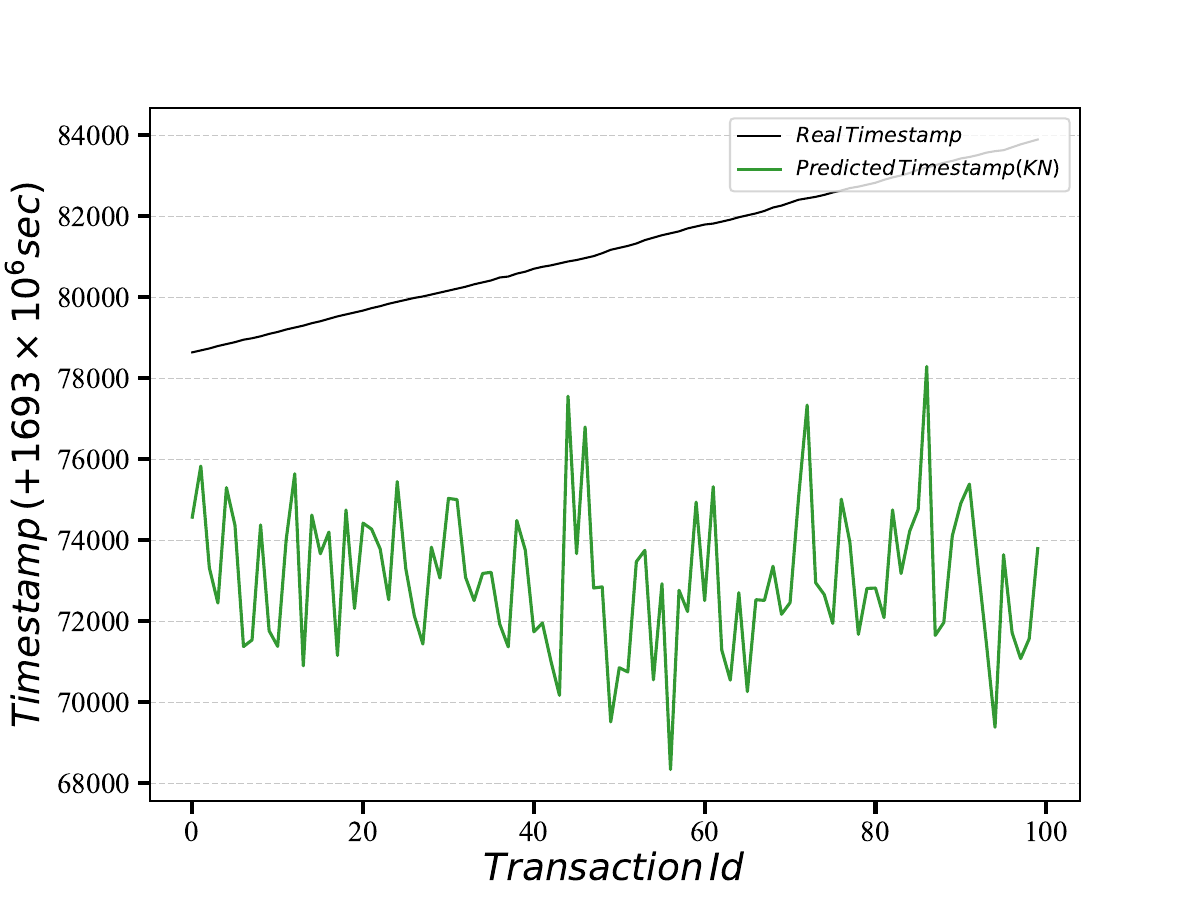}}
  \caption{Prediction results for transaction time}
  \label{fig:txnTime}
\end{figure}

\textbf{\textit{\underline{Transaction Time Prediction.}}} In the Ethereum network, \emph{txnFee} can significantly increase due to congestion and high demand for transaction confirmations. To mitigate the risk of higher costs, we propose a machine learning model to predict the optimal \emph{txnTime} when \emph{txnFee} is minimal. Similar to previous experiments, we plotted the line charts for the best and worst-performing models in predicting \emph{txntime} (see Fig.~\ref{fig:txnTime}). As with the previous charts, we focus on the systematic sample of 100 transactions for clarity. Each chart displays two lines—one indicating the actual timestamps (in black) and the other representing predicted timestamps (in green) from our machine-learning models. \emph{LR} performed very well with little deviation from the actual values (see Fig.~\ref{fig:txnTime}(a)). At the same time, the \emph{KN} exhibits a significant deviation from actual values (see Fig.~\ref{fig:txnTime}(b)). Due to space limitations, we have not included a performance comparison table for the models. This approach enables users to time their transactions strategically, leveraging the model's predictions to minimize \emph{txnFee} during periods of network congestion. Visualizing model performance through line charts effectively enhances our understanding of each model's predictive capabilities, emphasizing the importance of selecting the most reliable model for optimal results.

As the efficiency of machine learning models is intricately tied to the training dataset, any changes in the Ethereum network's dynamics can impact quantitative results while qualitative outcomes remain consistent. Significant shifts in Ethereum network conditions may necessitate modifications to machine learning models. For instance, a model yielding highly accurate results in the current scenario may fail to do so in a different scenario. Hence, a growing research direction aims to evolve machine learning models in the changing Ethereum network.

\section{Related Work}\label{rel_work}
The user determines the \emph{txnFee}; the transaction inclusion probability in the ledger increases, and the transaction wait time decreases with the increase in \emph{txnFee}; this has led to the need for solutions that can offer insights to users on setting transaction fees depending on their custom requirements (refer to Fig.~\ref{fig:usecase} for better clarity). Across the blockchain frameworks (e.g., Bitcoin~\cite{tsang2021market} and Ethereum~\cite{DBLP:journals/corr/abs-2012-00854, DBLP:conf/ccs/LiuLNZZZ22}), TFM is explored with a focus on user experience in transacting with such systems. We are restricting it to Ethereum as our work only aims to understand the miner dynamics and TFM in the context of Ethereum. 

Research on TFM has been prevalent among the Ethereum community since the EIP-1559 release~\cite{DBLP:journals/corr/abs-2012-00854, DBLP:conf/ccs/LiuLNZZZ22, chung2023foundations}. \emph{EIP-1559}~\cite{EIP-1559} proposes a major change to Ethereum's TFM and reforms the \emph{txnFee} market. It introduces a new fee structure involving \emph{baseFee} and \emph{priorityFee}. Users must pay a dynamic \emph{baseFee}, which adjusts based on network demand i.e., the \emph{baseFee} increases when the network is congested and decreases when the network is less busy. Additionally, they can pay a \emph{priorityFee} to expedite the inclusion of their transactions~\cite{DBLP:journals/corr/abs-2012-00854}. The \emph{baseFee} is burnt, thereby reducing the overall supply of Ethereum, while the \emph{priorityFee} is given to proposers.

\emph{EIP-1559} has improved the user experience by reducing the variation in gas prices within each block. This is achieved by fixing the \emph{baseFeePerGas} for all transactions in a block, thus reducing users' waiting time~\cite{DBLP:conf/ccs/LiuLNZZZ22}. One of the challenges associated with cryptocurrency transactions is determining the optimal \emph{txnFee}. If the fee is too low, the transaction may take a long time to process, while users risk overpaying if it is too high. Arnaud et al.~\cite{laurent2022transaction} have proposed a prediction and optimization method that combines the Monte Carlo approach~\cite{doi:10.1080/01621459.1949.10483310} and a binary search approach to address this problem. This method predicts the probability of transaction confirmation and determines the optimal \emph{txnFee} required for confirmation within a specified time limit. 

In Ethereum, \emph{txnFee} depends on gas price. Given the variability of factors, including difficulty, block gas limit, transaction gas limit, and ether price, it is crucial to determine the optimal gas price. Failing to do so can result in delayed transaction processing and higher costs, negatively impacting user experience and financial efficiency. Fangxiao et al.~\cite{DBLP:conf/dsa/LiuWLXG19} introduced a Machine Learning Regression-based approach (MLR) to address these challenges. MLR predicts the optimal gas price for the upcoming block, ensuring efficient transaction confirmation and cost minimization with an accuracy of 74.9


These works explore TFM for the \emph{PoW-Ethereum} scenario, which has become obsolete since Ethereum transitioned to PoS, changing the miners' dynamics. In this paper, we empirically analyze miner dynamics to assess its effect on decentralization and utilize regression-based machine learning models to predict \emph{txnFee} and \emph{txnTime} considering the \emph{PoS-Ethereum} scenario.


\section{Conclusion and Discussion}\label{conclusion}

We have comprehensively studied how Ethereum's \emph{Paris upgrade} (i.e., PoW to PoS) significantly impacts the miners' dynamics. Empirical research unveils findings (some anticipated and others unexpected) that are important for the Ethereum community. Anticipated findings are increased miner participation and lower \emph{txnFee}, which align with our intuitions. A surprising finding is that miner selection randomness is negatively affected after the Paris upgrade.

\emph{EIP-3675} substantially increased the unique miners ($\approx50\times$ PoW). Moreover, analyzing miners category-wise reveals a significant increase in the participation of \textit{\textbf{small-scale}} miners ($\approx60\times$ PoW). Conversely, the impact on the number of medium and large miners is negligible. This suggests that \emph{PoS-Ethereum} empowers \textit{\textbf{small-scale}} miners by enabling their participation without requiring significant computing resources, encouraging broader miner engagement in block proposals. However, \emph{EIP-3675} reduces the proposer selection randomness, which signifies this upgrade's negative impact on network decentralization. 

\emph{EIP-3675} also affects the TFM as validators have started considering transactions with lower fees, which led to lower but volatile \emph{txnFee}. Therefore, we utilized regression-based machine learning to predict \emph{txnFee} (i.e., \emph{baseFee} and \emph{priorityFee}), determine how much \emph{priorityFee} the user should pay so that their transaction gets confirmed without much delay, and predict the optimal transaction time on which the user should invoke the transaction so that the \emph{txnFee} is lesser. The experimental results showed that \emph{ET}, \emph{GB}, and \emph{LR} perform best in predicting \emph{baseFee}, \emph{priorityFee}, and \emph{txnFee}, respectively, while \emph{KN} performs worst. As the Ethereum network is encountering several significant upgrades, continual modifications to machine-learning models may be necessary. Hence, a growing research direction aims to evolve machine learning models in the Ethereum network. 





\bibliographystyle{unsrt} 
\renewcommand*{\bibfont}{\small}
\makeatletter
\def\url@leostyle{%
  \@ifundefined{selectfont}{\def\UrlFont{\sf}}{\def\UrlFont{\small\ttfamily}}}
\makeatother
\urlstyle{leo}
\bibliography{bibliograpghy}

\end{document}